\documentclass[aps,prfluids,superscriptaddress,longbibliography, floatfix]{revtex4}

\usepackage{CJK}
\usepackage{inputenc}
\usepackage{color}
\usepackage{graphicx}
\usepackage{float}
\usepackage[geometry]{ifsym}
\usepackage{subfigure}
\usepackage{amsmath, bm}
\def\3He{$^3$He} 
\def\4He{$^4$He}

\begin{document}
\title{An exploration of Homogeneous Isotropic Turbulence in He II Using Particle Tracking Velocimetry}

\author{Yuan Tang}
\affiliation{National High Magnetic Field Laboratory, 1800 East Paul Dirac Drive, Tallahassee, Florida 32310, USA}
\affiliation{Mechanical Engineering Department, Florida State University, Tallahassee, Florida 32310, USA}

\author{Shiran Bao}
\affiliation{National High Magnetic Field Laboratory, 1800 East Paul Dirac Drive, Tallahassee, Florida 32310, USA}
\affiliation{Mechanical Engineering Department, Florida State University, Tallahassee, Florida 32310, USA}

\author{Toshiaki Kanai}
\affiliation{National High Magnetic Field Laboratory, 1800 East Paul Dirac Drive, Tallahassee, Florida 32310, USA}
\affiliation{Department of Physics, Florida State University, Tallahassee, Florida 32306, USA}

\author{Wei Guo}
\email[Corresponding: ]{wguo@magnet.fsu.edu}
\affiliation{National High Magnetic Field Laboratory, 1800 East Paul Dirac Drive, Tallahassee, Florida 32310, USA}
\affiliation{Mechanical Engineering Department, Florida State University, Tallahassee, Florida 32310, USA}

\date{\today}

\begin{abstract}
Despite being a quantum two-fluid system, superfluid helium-4 (He II) is observed to behave similarly to classical fluids when a flow is generated by mechanical forcing. This similarity has brought up the feasibility of utilizing He II for high Reynolds number classical turbulence research, considering the small kinematic viscosity of He II. However, it has been suggested that the non-classical dissipation mechanism in He II at small scales may alter its turbulent statistics and intermittency. In this work, we report our study of a nearly homogeneous and isotropic turbulence (HIT) generated by a towed grid in He II. We measure the velocity field using particle tracking velocimetry with solidified deuterium particles as the tracers. By correlating the velocities measured simultaneously on different particle trajectories or at different times along the same particle trajectory, we are able to conduct both Eulerian and Lagrangian flow analyses. Spatial velocity structure functions obtained through the Eulerian analysis show scaling behaviors in the inertial subrange similar to that for classical HIT but with enhanced intermittency. The Lagrangian analysis allows us to examine the flow statistics down to below the dissipation length scale. Interestingly, strong deviations from the classical scaling behaviors are observed in this regime. We discuss how these deviations may relate to the motion of quantized vortices in the superfluid component in He II.
\end{abstract}
\maketitle

\section{Introduction} \label{SecI}
Below about 2.17 K, liquid $^4$He undergoes a second-order phase transition to the superfluid phase (He II), which consists of two fully miscible fluid components: an inviscid superfluid component (i.e., the condensate) and a viscous normal-fluid component (i.e., the thermal excitations) \cite{Tilley-book}. The rotational motion in the superfluid can occur only with the formation of topological defects in the form of quantized vortex lines \cite{Donnelly-book}. These vortex lines all have identical cores (thickness $\xi_0\simeq1$~{\AA}) and they each carry a single quantum of circulation $\kappa\simeq10^{-3}$ cm/s. Turbulence in the superfluid takes the form of an irregular tangle of vortex lines (quantum turbulence) \cite{Vinen-2002-JLTP}. The normal fluid behaves more like a classical fluid. But a force of mutual friction between the two fluids \cite{Vinen-1957-PRS}, arising from the scattering of thermal excitations by the vortex lines, can affect the flows in both fluids.

Despite being a quantum two-fluid system, He II has been observed to exhibit flow characteristics similar to that in classical fluids when the flows are generated by mechanical forcing \cite{Stalp-1999-PRL,Maurer-1998-EPL}. This similarity has brought up the feasibility of utilizing the small kinematic viscosity of He II (i.e., about three orders of magnitudes smaller than that of ambient air \cite{Donnelly-1998-JPCRD}) to generate turbulent flows with extremely high Reynolds numbers for classical turbulence research and model testing \cite{Sreenivasan-2001-AAM, Niemela-2006-JLTP}. The quasiclassical behavior of He II in mechanically driven flows is believed to be the result of a strong coupling of the two fluids at large scales by mutual friction \cite{Vinen-2000-PRB}. The turbulent eddies in the normal fluid are matched by eddies in the superfluid induced by local polarization of the vortex tangle \cite{Barenghi-1997-PF}. However, at small scales, especially below the mean inter-vortex distance $\ell$, this coupling must break down because the superfluid flow is then controlled by the discrete vortex lines and cannot match the classical normal-fluid flow. Therefore, a mutual friction dissipation sets in at these small scales, in addition to the viscous dissipation in the normal fluid \cite{Vinen-2002-JLTP}. This unique small-scale dissipation mechanism in He II can give rise to subtle differences between He II quasiclassical flows and flows in classical fluids. For instance, a past theoretical work suggested a temperature-dependent enhancement of turbulence intermittency in He II quasiclassical flows \cite{Boue-2013-PRL}. In order to explore these interesting similarities and differences, quantitative velocity-field measurements in a simple and well-controlled He II quasiclassical flow are needed.

A simple form of turbulence that has received extensive attention in classical fluids research is the so-called homogeneous isotropic turbulence (HIT) \cite{Comte-1966-JFM,Tennekes-1972-B,Sinhuber-2015-PRL}, which can also be achieved in He II in the wake of a towed grid \cite{Stalp-1999-PRL}. In a recent work, we reported the study of grid turbulence in a He II filled channel using a molecular tagging velocimetry (MTV) technique \cite{Varga-2018-PRF}. This technique is based on the creation and tracking of thin lines of He$_2^*$ molecular tracers \cite{Gao-2015-RSI}. These tracers are completely entrained by the viscous normal fluid above 1 K and therefore their motion provides unambiguous information about the normal-fluid flow \cite{Marakov-2015-PRB, Gao-2016-JETP, Gao-2016-PRB, Gao-2017-PRB, Gao-2017-JLTP, Gao-2018-PRB, Bao-2018-PRB}. A striking nonmonotonic temperature-dependent intermittency enhancement was observed for the first time \cite{Varga-2018-PRF}. Nevertheless, there are two major limitations in the MTV experiment: 1) the MTV method only allows the measurement of the velocity component perpendicular to the tracer lines \cite{Guo-2019-JLTP}, and hence there lacks information about the isotropicity of the flow; and 2) the spatial resolution is limited by the displacement of the tracer lines (i.e., 100-200 $\mu$m), which is greater than the typical dissipation length scale (i.e., a few tens of microns).

To overcome these issues, we report in the present work the application of a particle tracking velocimetry (PTV) technique for velocity-field measurements in a recently built He II grid turbulence facility \cite{Mastracci-2018-RSI}. Micron-sized solidified deuterium particles are used as the tracers, whose motion can be tracked with a spatial resolution of a few microns to render both the horizontal and the vertical velocities within the imaging plane. It is known that these relatively large particles can get trapped on quantized vortices in He II besides experiencing the drag force from the normal fluid \cite{Mastracci-2017-JLTP}, which makes it challenging to interpret their motion in He II flows where the two fluids have different mean velocities (e.g., heat-induced thermal counterflow \cite{Landau-book}) \cite{Mastracci-2018-PRF, Mastracci-2019-PRF,Mastracci-2019-PRF-2}. However, this issue becomes an advantageous feature in grid turbulence. At large scales where the two fluids are coupled, the motion of the particles can provide us quantitative information about the coupled velocity field. At small scales, deviations from the classical turbulence statistics may be unveiled due to the motion of the trapped particles.

In Sec.~\ref{SecII}, we briefly describe the experimental setup and the measurement methods. In Sec.~\ref{SecIII}, we first present evidences to show that a nearly HIT can emerge in the decay of the towed-grid generated turbulence in He II. Then, by correlating the velocities measured simultaneously on different particle trajectories or at different times along the same particle trajectory, we manage to conduct both Eulerian and Lagrangian velocity statistical analyses. We show that the spatial velocity structure functions obtained through the Eulerian analysis exhibit scaling behaviors in the inertial subrange similar to that for classical HIT but with enhanced intermittency. The Lagrangian analysis, on the other hand, allows us to examine the flow statistics down to below the dissipation length scale. In this regime, strong deviations from the classical scaling behaviors are observed. We discuss how these deviations may relate to the motion of the quantized vortices. A brief summary is provided in Sec.~\ref{SecIV}.

\section{Experimental Method} \label{SecII}
\begin{figure}[htb]
  \centering
  \includegraphics[width=0.55\linewidth]{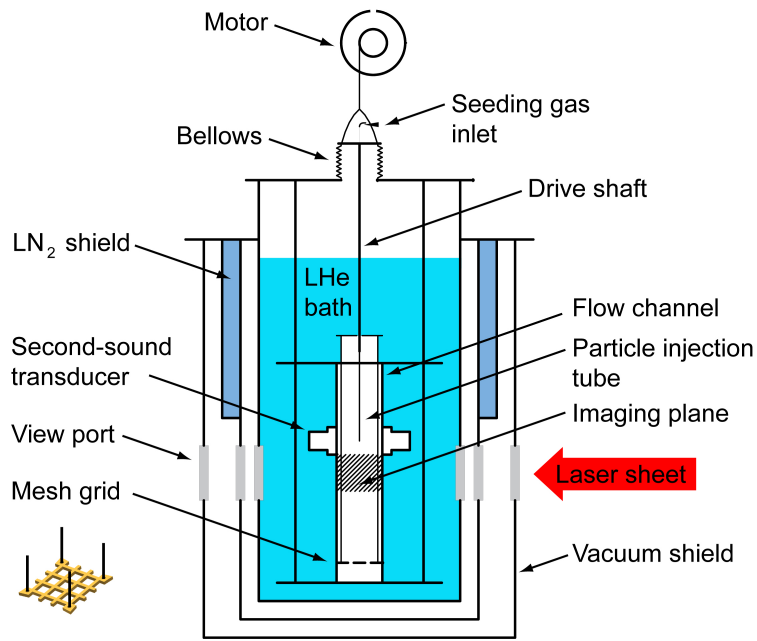}
  \caption{Schematic diagram of the experimental apparatus. }
  \label{Fig:apparatus}
\end{figure}

The experimental apparatus used in the current work was designed and built specifically for PTV-based He II grid-turbulence research \cite{Mastracci-2018-RSI}. As shown schematically in Fig.~\ref{Fig:apparatus}, a transparent cast acrylic flow channel (cross-section area: 1.6$\times$1.6 cm$^2$, length: 33 cm) is immersed vertically in a He II bath, where the helium temperature can be controlled by regulating the vapor pressure in the bath. A brass mesh grid is suspended by four stainless-steel thin wires at the four corners inside the flow channel. These wires are connected to the drive shaft of a linear motor system. A LabVIEW computer program is developed to control the motor system such that the grid can be pulled at a constant speed between 0.1 and 60 cm/s. In this specific work, we use a fixed grid speed at 30 cm/s. In order to minimize possible large-scale secondary flows around the moving grid, we followed the guidelines from classical grid turbulence research \cite{Fernando-1993-PoF,Honey-2014-PRE} and designed our grid to have a mesh spacing of 3 mm and an open area of 40\% and with special treatments of its boundary and the four corners \cite{Mastracci-2018-RSI}.

To probe the flow, solidified deuterium particles are used as the tracers. These particles are produced by slowly injecting a room-temperature mixture of 5\% deuterium gas and 95\% helium gas through a leak valve into the flow channel \cite{Fonda-2016-RSI}. The gas mass flow rate is restricted such that the injection does not affect the bath temperature. Typically, about 70\% of the resulting particles have diameters in the range 3 to 6 $\mu$m, as determined from their settling velocity in quiescent He II \cite{Mastracci-2018-RSI}. A continuous-wave laser sheet (thickness: 200 $\mu$m, height: 9 mm) passes through the geometric center of the channel to illuminate the particles. We then pull the grid at the controlled speed and use a high-speed camera (120 frame per second) to record the motion of the particles. Due to the camera's limited internal memory, we record the particle motion for a period of 0.28 s (i.e., 34 frames) for every 2 s following the passage of the grid. Particle trajectories can be extracted from the sequence of images based on the feature-point tracking routine developed by Sbalzarini and Koumoutsakos \cite{Sbalzarini-2005-JSB}.

Besides the PTV measurement, a standard second-sound attenuation method is also used to measure the temporal evolution of the spatial-averaged vortex-line density $L(t)$ (i.e., total length of the vortices per unit volume, $L^{-1/2}=\ell$) \cite{Mastracci-2018-RSI}. The turbulence generated by the towed grid decays with time. We take the instant when the grid passes through the center of the view port as the time origin for both the visualization and the second-sound measurements. These measurements are made in the temperature range of 1.65 to 2.12 K. At each temperature, we normally repeat the experiment 10 times so that an ensemble statistical analysis of the particle trajectories can be performed at different decay times.

\section{Experimental Results and discussions} \label{SecIII}
\subsection{Temporal evolution of the grid turbulence}
\begin{figure}[b]
\centering
\includegraphics[width=0.9\textwidth]{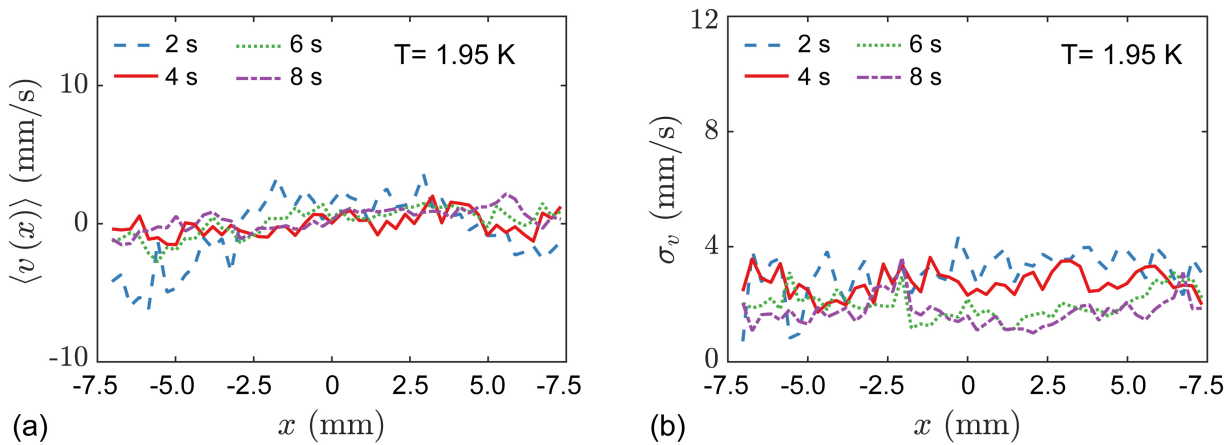}
\caption{(a) Ensemble-averaged vertical velocity $\left\langle v(x)\right\rangle$ profile at various decay times following the passage of the grid. (b) The corresponding vertical velocity variance $\sigma_v$. The data were taken at 1.95 K with a grid velocity of 30 cm/s.}
\label{Fig2}
\end{figure}
It is generally believed that a moving grid in He II first produces turbulent eddies with sizes comparable to the mesh grid spacing. Then, after a short transient period, the energy-containing eddies saturate at sizes comparable to the width of the channel, which leads to a nearly HIT that decays with time \cite{Stalp-1999-PRL}. However, the observation of large-scale eddies right after the passage of the grid in our previous MTV experiment casts doubt on this simple physical picture \cite{Varga-2018-PRF}. These large-scale eddies are likely due to the secondary flows caused by the imperfection of the grid geometry, which is hard to completely avoid. To examine the evolution of the velocity profile in our current experiment, we analyze the particle trajectories and calculate the vertical velocity $v(x)$ as a function of the horizontal position $x$ across the width of the flow channel. Representative velocity profiles $\left\langle v(x)\right\rangle$ obtained at 1.95 K through an assemble average over many trajectories and over 10 experimental trials are shown in Fig.~\ref{Fig2}~(a). It is obvious that large-scale flows do exist at short decay times, despite the careful design of the grid. Nevertheless, these eddies have all decayed by $t=4$ s such that the mean velocity $\left\langle v(x)\right\rangle$ is nearly zero across the entire channel width. Fig.~\ref{Fig2}~(b) shows the profile of the corresponding vertical-velocity variance $\sigma_v$, defined as $\sigma_v=\left\langle[v(x)-\overline{v}]^2\right\rangle^{1/2}$. The variance $\sigma_v$ appears to be more spatially homogeneous and remains at a relatively high level at $t=4$ s.

\begin{figure}[h]
\centering
\includegraphics[width=0.9\textwidth]{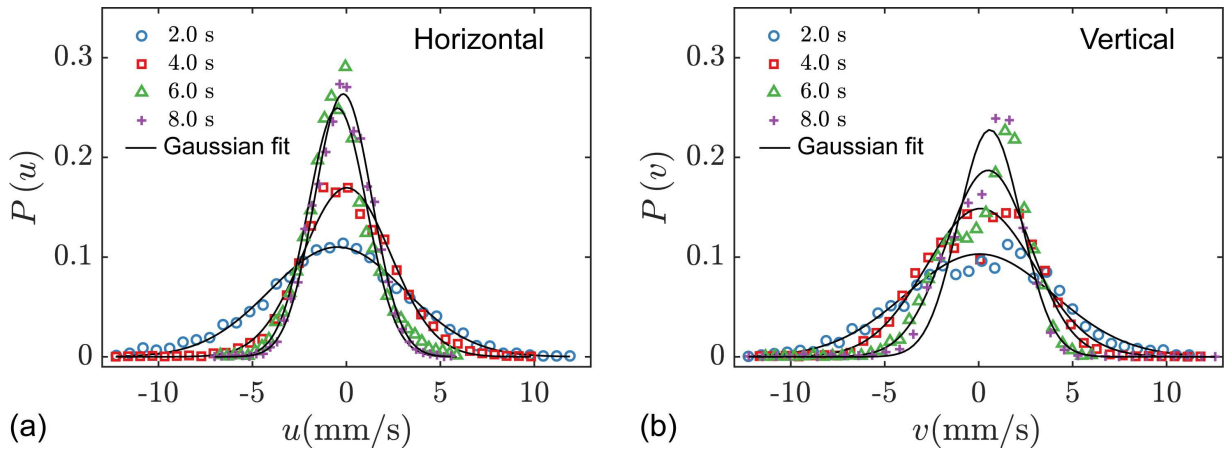}
\caption{Horizontal (a) and Vertical (b) particle velocity PDFs at different decay times as indicated. The data were taken at 1.95 K with a grid velocity of 30 cm/s.}
\label{Fig3}
\end{figure}

In Fig. \ref{Fig3}, we show the calculated probability density functions (PDFs) of the particle horizontal velocity $u$ and the vertical velocity $v$ obtained at 1.95 K. At small decay times, the vertical velocity PDFs exhibit double-peak structures, which reflect the large-scale eddies as revealed in Fig.~\ref{Fig2}~(a). After these large-scale eddies decay, the velocity PDFs can be fitted reasonably well by Gaussian functions. Through such fits, the evolution of the velocity variances in both the horizontal direction $\sigma_u(t)$ and the vertical direction $\sigma_v(t)$ can be obtained, which provides us information about the decay of the turbulence kinetic energy density $K_{u}(t)=\frac{1}{2}{\sigma_u(t)}^2$ and $K_{v}(t)=\frac{1}{2}{\sigma_v(t)}^2$.

\begin{figure}[h]
\centering
\includegraphics[width=0.9\textwidth]{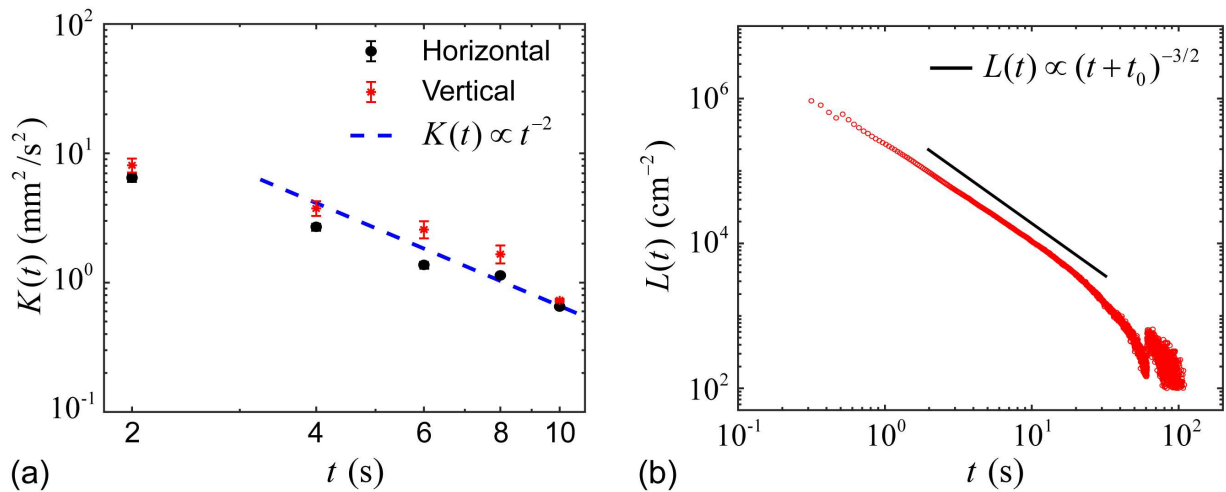}
\caption{(a) Time evolution of the turbulence kinetic energy $K(t)$ contribution from the horizontal and the vertical velocity components. (b) Decay of the quantized vortex-line density $L(t)$. The solid black curve represents the scaling $L(t)\propto(t+t_0)^{-3/2}$, where $t_0=0.27$ s is the virtual time origin \cite{Skrbek-2000-PoF}. The data were taken at 1.95 K.}
\label{Fig4}
\end{figure}

In Fig. \ref{Fig4}, we show the time-evolution of the obtained turbulence kinetic energy $K(t)$ together with the measured vortex-line density $L(t)$. It turns out that $K(t)$ decays more or less accordingly to $K(t)\propto t^{-2}$, especially for $t\geq4$ s. The contributions to $K(t)$ from the two velocity components appear to have similar magnitudes, which suggests that the turbulence is relatively isotropic. The decay of the vortex-line density exhibits a scaling behavior of $L(t)\propto{t^{-3/2}}$ after the first a few seconds. Both these scalings are considered as the characteristics of decaying HIT in He II when the sizes of the energy-containing eddies are saturated by the channel width \cite{Vinen-2000-PRB,Stalp-1999-PRL}.

Based on the above analysis, the turbulence at 4 s decay time appears to be reasonably homogeneous and isotropic, and its turbulence kinetic energy density is relatively high such that an inertial subrange may exist. In what follows, we shall focus on the data set taken at $t=4$ s for detailed statistical analysis.

\subsection{Eulerian structure functions and intermittency}
In the Eulerian description of fluid flow, spatial structure functions are known to be very useful tools for characterizing the statistical properties of the turbulence \cite{Davidson-2004-book}. For fully developed HIT, the relevant forms of the structure functions are the $n$-th order longitudinal and transverse structure functions, defined as \cite{Stolovitzky-1993-PRE,Benzi-1995-PhysicaD}:
\begin{equation}
  S_n^\|(r) = \left\langle\left|\delta \mathbf{V}(\mathbf{r})\cdot{\hat{\mathbf{r}}}\right|^n\right\rangle
~~\mathrm{and}~~
  S_n^\bot(r) = \left\langle\left|\delta \mathbf{V}(\mathbf{r})\times{\hat{\mathbf{r}}}\right|^n\right\rangle
\label{eq:Sn}
\end{equation}
where $\delta \mathbf{V}(\mathbf{r})=\mathbf{V}(\mathbf{r_1})-\mathbf{V}(\mathbf{r_2})$ denotes the difference of the velocities measured simultaneously at two locations that are separated by $\mathbf{r}=\mathbf{r_1}-\mathbf{r_2}$, and the angle brackets represent the ensemble average. For fully-developed ideal HIT in an incompressible fluid, these structure functions exhibit the well-known Kolmogorov-Obukhov scaling behaviors \cite{Frisch-1995-B}. Specifically, the third-order structure function should scale as $|S_3(r)|=\frac{4}{5}\epsilon r$, where $\epsilon=-\frac{dK}{dt}$ is the energy dissipation rate. The range of $r$ over which this scaling holds defines the inertial subrange of the turbulence energy cascade. In this inertial subrange, the second-order structure function is expected to scale as $S_2(r)\propto r^{2/3}$.

\begin{figure}[b]
\centering
\includegraphics[width=0.88\textwidth]{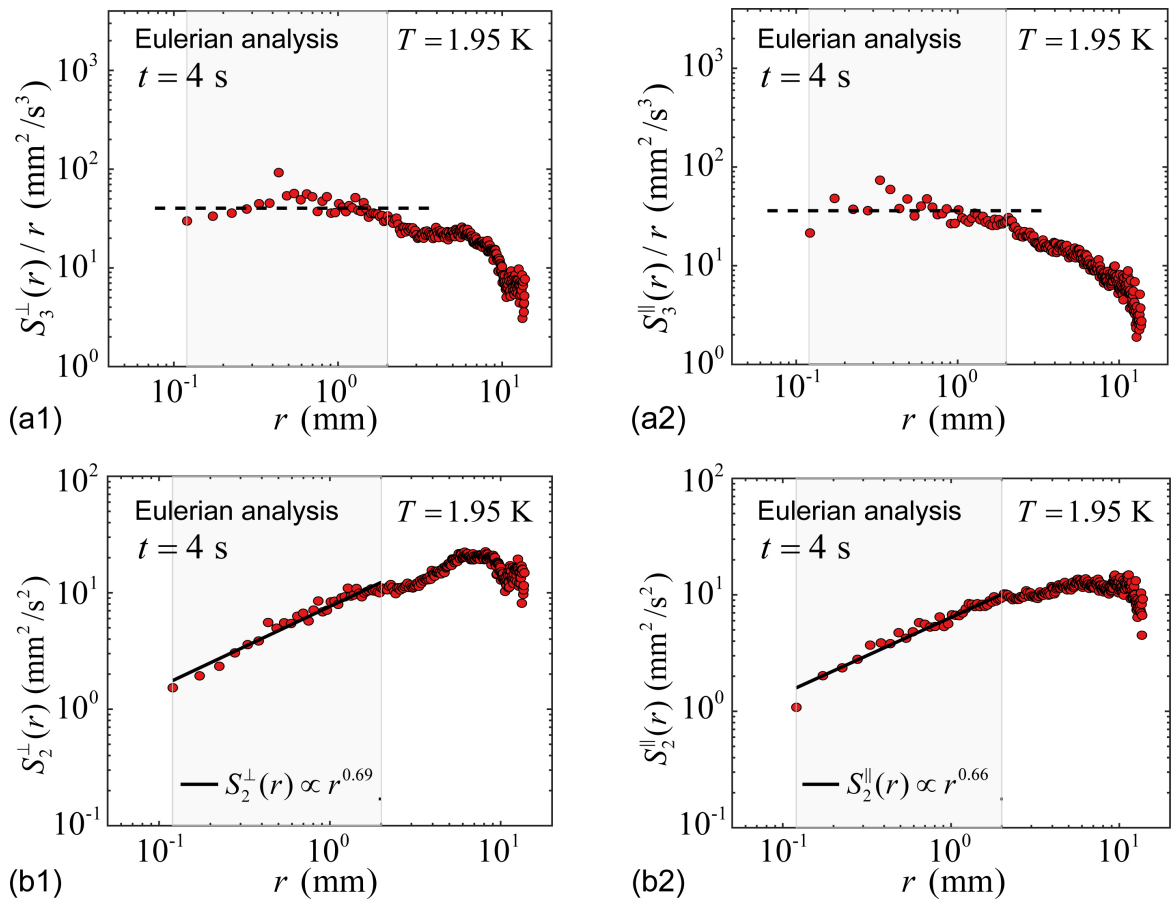}
\caption{(a1) and (a2) show the third-order transverse and longitudinal velocity structure functions compensated by $r^{-1}$. The dashed horizontal lines are drawn to guide the eye. The shaded region indicates the inertial subrange. (b1) and (b2) show the second-order velocity structure functions. The solid lines are power-law fits to the data in the shaded region.}
\label{Fig5}
\end{figure}

To check whether an inertial subrange develops in our grid turbulence, we have performed Eulerian analysis of the velocity field by correlating the velocities measured simultaneously on different particle trajectories. To our knowledge, no prior PTV experiments with He II have ever reported the implementation of the Eulerian analysis. In Fig.~\ref{Fig5} (a1) and (a2), we show the calculated $S_3^\|(r)$ and $S_3^\bot(r)$ compensated by $r^{-1}$ for the data set obtained at $t=4$ s. Over the range 0.12~mm$\leq r\leq$2~mm as highlighted by the shaded region, both $S_3^\|(r)/r$ and $S_3^\bot(r)/r$ appear to be more or less flat, indicating the existence of a cascade inertial subrange. We also plot the second-order structure functions in Fig.~\ref{Fig5} (b1) and (b2). It is clear that in the inertial subrange, both $S_2^\|(r)$ and $S_2^\bot(r)$ can be well fitted by power-law scalings that are indeed very close to $r^{2/3}$. Note that the range of $r$ in the Eulerian analysis are set by our requirement that the sample number at a given particle separation $r$ is greater than $10^2$ (see Appendix~\ref{App-A} for more details).

\begin{figure}[h]
\centering
\includegraphics[width=0.85\textwidth]{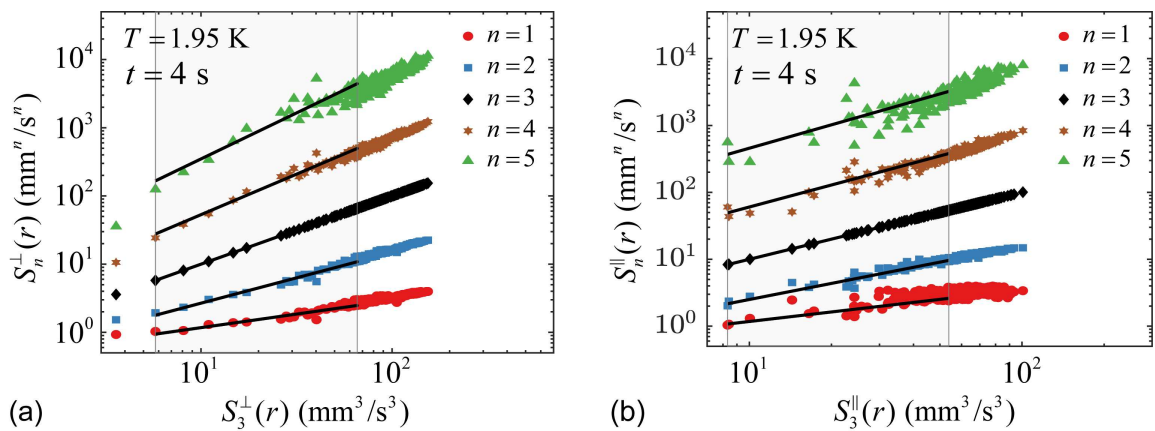}
\caption{Extended self-similarity plots of (a) transverse and (b) longitudinal velocity structure functions for $n=1\sim5$ versus the third-order structure functions. The solid lines are power-law fits to the data in the inertial subrange.}
\label{Fig6}
\end{figure}

The Kolmogorov-Obukhov scalings of the higher-order structure functions in the inertial subrange in an ideal HIT are $S_n(r) \propto r^{\frac{n}{3}}$ \cite{Hinze-1975-B}. However, intermittency can occur spontaneously in real turbulent flows, which manifests itself as extreme velocity excursions that appear more frequently than one would expect on the basis of Gaussian statistics. Corrections to the scaling exponents of the velocity structure functions are therefore expected, especially for higher-order structure functions that are more sensitive to the occurrence of rare events. She and Leveque proposed a universal scaling $S_n(r)\propto r^{\zeta_n}$ for HIT in classical fluids \cite{She-1994-PRL}, where $\zeta_n = \frac{n}{9}+2\left[1-(\frac{2}{3})^{n/3}\right]$. These predicted scalings were confirmed experimentally by Benzi \emph{et al.} \cite{Benzi-1993-PRE}. To examine the scaling behaviors of the structure functions and the intermittency in He II grid turbulence, we adopt the extended self-similarity (ESS) method by plotting $S_n(r) $ versus $S_3(r)$ (instead of $r$) in the inertial subrange \cite{Benzi-1993-PRE,Salort-2011-JoP}. It is known that the ESS analysis can reveal scaling laws even for turbulent flows with moderate Reynolds numbers \cite{Benzi-1993-EPL}, thereby allowing for more accurate determination of the scaling exponents \cite{Dubrulle-1994-PRL}. In Fig.~\ref{Fig6}, we show the calculated $S_n^\|(r)$ and $S_n^\bot(r)$ versus $S_3(r)$. Clear power-law dependance of $S_n^\|(r)$ and $S_n^\bot(r)$ on $S_3(r)$ that extend beyond the inertial subrange are observed.

\begin{figure}[h]
\centering
\includegraphics[width=0.85\textwidth]{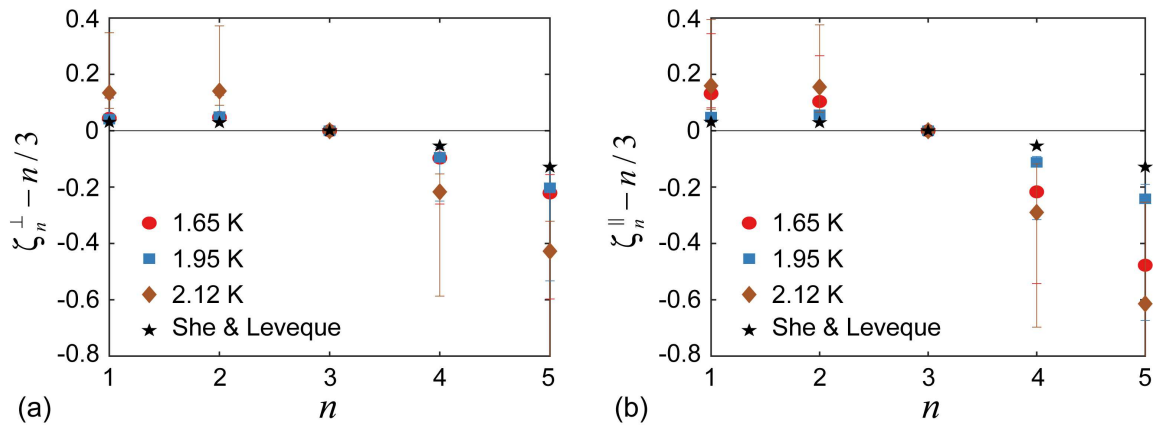}
\caption{Intermittency corrections to the scaling exponents of the transverse (a) and longitudinal (b) structure functions for He II grid turbulence at various temperatures. The corrections for classical fluids are also included for comparison \cite{She-1994-PRL}.}
\label{Fig7}
\end{figure}
In Fig.~\ref{Fig7}, we show the scaling exponents $\zeta_n$ extracted through the power-law fits in the ESS analysis. The large error bars are due to the relatively strong scattering of the data points in the fits. Besides the data obtained at 1.95 K (48.2\% normal fluid), the results of similar measurements conducted at 1.65 K (19.3\% normal fluid) and 2.12 K (78.8\% normal fluid), together with the $\zeta_n$ values that She and Leveque proposed for classical fluids \cite{She-1994-PRL}, are also collected in Fig.~\ref{Fig7}. The differences between $\zeta_n$ and the Kolmogorov-Obukhov scalings of $n/3$ are clearly seen in our data, which confirms the existence of intermittency in He II grid turbulence. The observation that $\zeta_n$ is universally smaller than $\frac{n}{3}$ for $n>3$ agrees well with the $\zeta_n$ behavior in classical fluids \cite{Stolovitzky-1993-PRE1}. Furthermore, it is clear that the intermittency in He II grid turbulence for $n>3$ is enhanced compared to that in classical fluids, which agrees with theoretical predictions \cite{Boue-2013-PRL}. However, due to the large error bars associated with the extracted $\zeta_n$, we cannot draw any definite conclusion regarding the temperature dependance of the intermittency in He II quasiclassical turbulence \cite{Varga-2018-PRF,Salort-2011-JoP,Rusaouen-2017-PoF}.


\subsection{Lagrangian analysis at small length scales}
We have also conducted Lagrangian analysis of the particle motion by correlating the velocities measured along individual particle trajectories at different times. Conventionally, temporal structure functions in the Lagrangian framework can be constructed as:
\begin{equation}
  S_n^\|(\tau) = \left\langle\left|\delta \mathbf{V}(\tau)\cdot{\hat{\mathbf{r}}}\right|^n\right\rangle
~~\mathrm{and}~~
  S_n^\bot(\tau) = \left\langle\left|\delta \mathbf{V}(\tau)\times{\hat{\mathbf{r}}}\right|^n\right\rangle
\label{eq:Sn}
\end{equation}
where $\delta \mathbf{V}(\tau)=\mathbf{V}(t+\tau)-\mathbf{V}(t)$ denotes the difference of the velocities measured at $t+\tau$ and $t$ along a single particle trajectory, and $\mathbf{r}$ is the displacement of the particle over the time interval $\tau$. In order to make more direct comparison with the Eulerian structure functions, in what follows, we will calculate the Lagrangian structure functions and plot them as a function of the distance $r$ instead of $\tau$. This treatment allows us to examine the flow statistics down to scales as small as the particle displacement in one frame time (i.e., 8.3 ms).

\begin{figure}[h]
\centering
\includegraphics[width=0.9\textwidth]{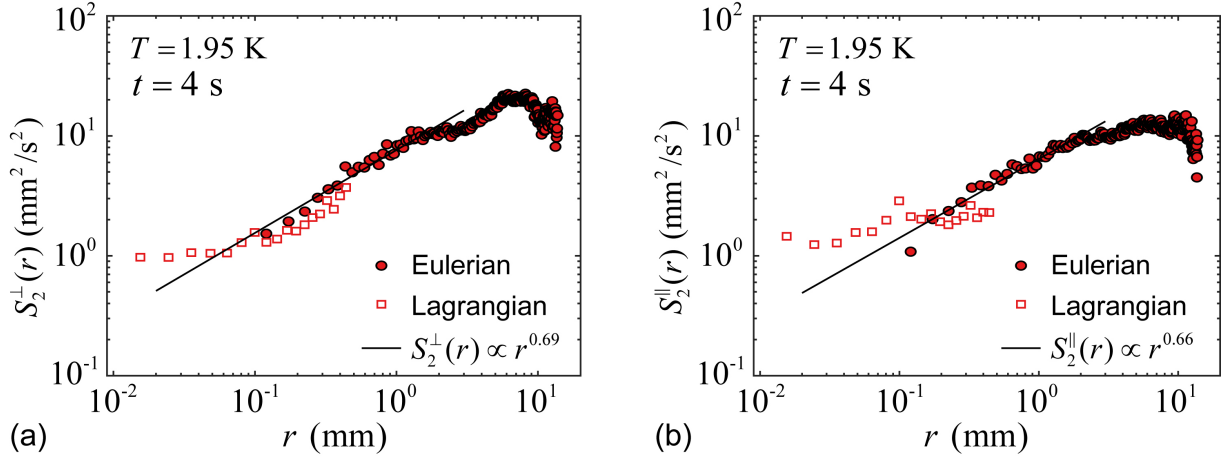}
\caption{Second-order (a) transverse and (b) longitudinal velocity structure functions obtained through both the Lagrangian and the Eulerian analyses of the particle trajectories for the data set obtained at 1.95 K and $t=4$ s.}
\label{Fig8}
\end{figure}

Fig. \ref{Fig8} shows the calculated second-order Lagrangian structure functions for the representative case of the He II grid turbulence at 1.95 K and at the decay time $t=4$~s. The range of $r$ covered in the Lagrangian analysis overlaps partly with that in the previous Eulerian analysis, while the lower bound of $r$ now extends down to about 15 $\mu$m, which is much smaller than the mean inter-vortex distance $\ell$ (i.e., about 54 $\mu$m based on Fig. \ref{Fig4}~(b)). Interestingly, in the overlapping region of $r$, the Lagrangian structure function data appear to agree quite well with that of the Eulerian analysis. This suggests that despite the different physical bases for the calculations of the Lagrangian and the Eulerian structure functions, they both exhibit similar scalings and magnitudes in the inertial subrange.

The more striking feature as revealed in Fig.~\ref{Fig8} is the deviation of the Lagrangian structure functions from the inertial-subrange scaling at length scales below about 50 $\mu$m, i.e., comparable to the mean inter-vortex distance $\ell$. Similar deviations are also observed at other temperatures. Indeed, the appearance of the deviation is not too surprising, because we know the energy dissipation must set in at length scales comparable to $\ell$ \cite{Gao-2018-PRB}, which terminates the inertial energy cascade. However, in classical turbulence, it has been known that the viscous dissipation leads to an asymptotic scaling of the second-order structure function as $S_2(r)\propto r^2$ at small scales  \cite{Hinze-1975-B, Stolovitzky-1993-PRE1}. This means that if the grid turbulence in He II truly behaves classically, one would see the $S_2(r)$ value drop rapidly in the dissipation subrange instead of rising above the inertial-subrange scaling curve. Therefore, an outstanding question is what causes the observed abnormal behavior of $S_2(r)$ at small scales.

To provide our thoughts on this question, let us consider what the tracer particles actually trace in He II grid turbulence. Note that these micron-sized particles can either get trapped on quantized vortices in the superfluid or entrained by the viscous normal fluid \cite{Mastracci-2019-PRF-2,Bewley-2006-Nature,Zhang-2005-NPhys}. At length scales much greater than $\ell$, the two fluids are strongly coupled by the mutual friction. Therefore, regardless whether the particles are trapped or not, their motion at large scales simply provides information about the coupled velocity field. At small scales where the normal-fluid motion is strongly damped by viscosity and the mutual friction \cite{Gao-2018-PRB}, the particles entrained by the normal fluid would make little contributions to the ensemble-averaging calculation of $S_2(r)$. On the other hand, for those trapped particles, their motions at small scales are controlled by the dynamics of individual quantized vortices. Even at scales below $\ell$, the vortices still move randomly with a characteristic mean velocity $\langle v^2_L\rangle^{1/2}$ given by \cite{Vinen-2002-JLTP,Gao-2018-PRB}:
\begin{equation}
\langle v^2_L\rangle^{1/2}=\frac{\kappa}{4\pi}\left\langle\frac{1}{R^2}\ln^2\left(\frac{R}{\xi_0}\right)\right\rangle^{1/2},
\end{equation}
where $R$ is the local curvature radius of the vortices. Therefore, the trapped particles can lead to appreciable values of $S_2(r)$ at small scales. The exact behavior of $S_2(r)$ in the dissipation subrange will then depend on the fraction of the particles that are trapped and the temporal velocity correlations of the vortices. We would like to point out that moving vortices can generate wake structures in the normal fluid due to the mutual friction \cite{Idowu-2000-PRB, Mastracci-2019-PRF-2, Yui-2020-PRL}. If an untrapped particle moves through such wake structures, it may experience velocity variations which lead to a finite contribution to $S_2(r)$ in the dissipation subrange. However, due to the small sizes of the wake structures, the dominant contribution to $S_2(r)$ should still come from the trapped particles. To test this physical picture, numerical simulations that can track the particles coupled to both the viscous normal fluid and the quantized vortices are needed \cite{Kivotides-2008-PRB,Kivotides-2008-PRB-2}, which is beyond the scope of this work.

\begin{figure}[h]
\centering
\includegraphics[width=0.9\textwidth]{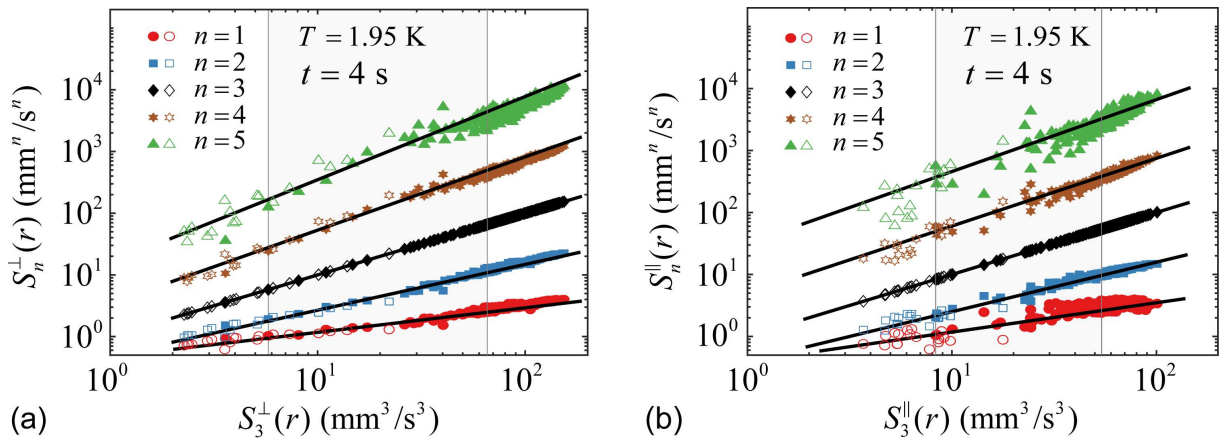}
\caption{Extended self-similarity plots of (a) transverse and (b) longitudinal velocity structure functions based on both Lagrangian analysis (empty symbols) and Eulerian analysis (solid symbols) of the particle trajectories for the data set obtained at 1.95 K and $t=4$ s. The solid lines represent the power-law fits to the Eulerian data as shown in Fig~\ref{Fig6}.}
\label{Fig9}
\end{figure}

Finally, we perform the ESS analysis of the Lagrangian structure functions with $n=1\sim5$ and plot them together with the Eulerian structure functions in Fig.~\ref{Fig9}. The Lagrangian structure functions extend to regions with smaller values of $S_3(r)$. Interestingly, despite the fact that the Lagrangian structure function data largely fall in the dissipation subrange, they appear to follow nicely the power-law scalings of the Eulerian data in the inertial subrange. This observation confirms the conclusion from classical turbulence research that the ESS scalings can encompass both the inertial and the dissipation subranges \cite{Arneodo-1996-EPL, Benzi-1993-PRE}.

\section{Summary} \label{SecIV}
We have conducted PTV study of a nearly HIT in He II which emerges in the decay of the turbulent flow produced by a towed grid in a flow channel. By correlating the velocities measured simultaneously on different particle trajectories or at different times along the same particle trajectory, we have conducted both Eulerian and Lagrangian analyses of the turbulent velocity field. We find that the spatial velocity structure functions obtained through the Eulerian analysis exhibit scaling behaviors in the inertial subrange similar to that for classical HIT but the intermittency is obviously enhanced. The Lagrangian analysis allows us to obtain information about the velocity field in both the inertial subrange and the dissipation subrange. In the inertial subrange, the Lagrangian structure functions show similar magnitudes and scaling behaviors as the Eulerian counterparts. However, they deviate strongly from the classical scalings in the dissipation subrange. We propose that this abnormal behavior is related to the tracer particles which are trapped on quantized vortices, the verification of which requires numerical simulations that account for the coupling of the particles to both the normal fluid and the quantized vortices.

\appendix
\section{Sample number distribution in the Eulerian and Lagrangian analyses}~\label{App-A}
\begin{figure}[h]
\centering
\includegraphics[width=0.9\textwidth]{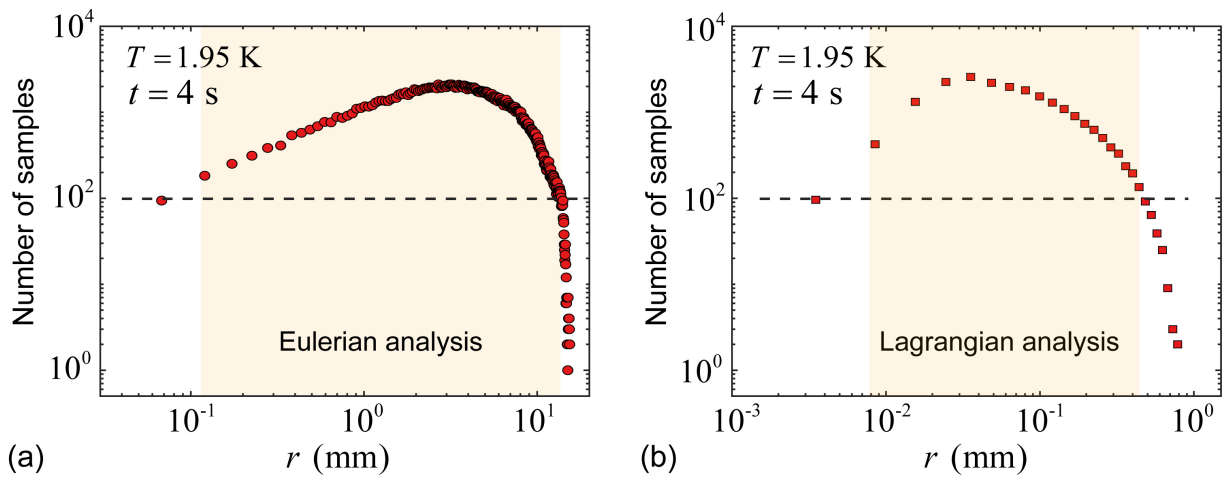}
\caption{Sample number as a function of $r$ for (a) Eulerian and (b) Lagrangian velocity statistical analyses of the particle trajectories for the data set obtained at 1.95 K and $t=4$ s. The analyses are conducted only in the shaded regions where the sample number is greater than $10^2$.}
\label{Fig10}
\end{figure}

The Eulerian velocity structure function analysis is conducted by correlating the velocities measured simultaneously on different particle trajectories. The range of $r$ covered in this analysis is limited by the minimum and the maximum separation distances between the particle pairs. The Lagrangian structure function analysis is based on correlating the velocities measured along individual particle trajectories at different times. As discussed in the text, the Lagrangian structure functions are plotted as a function of the particle displacement $r$ instead of the drift time. The corresponding range of $r$ is then limited by the minimum and the maximum displacement of individual particles. In Fig. \ref{Fig10}, we show the sample numbers extracted from a representative data set obtained at 1.95 K and at $t=4$ s as a function of $r$. In order for improved accuracy in the ensemble-averaging calculations of the structure functions, we only analyze the data in the region of $r$ where the sample number is greater than $10^2$.

\begin{acknowledgments}
This work is supported by the National Science Foundation (NSF) under Grant No. DMR-1807291 and partially by U.S. Department of Energy under Grant No. DE-SC0020113. The experiment was conducted at the National High Magnetic Field Laboratory at Florida State University, which is supported through the NSF Cooperative Agreement No. DMR-1644779 and the state of Florida.
\end{acknowledgments}

\bibliography{PTV-arxiv}

\begin{thebibliography}{55}
\expandafter\ifx\csname natexlab\endcsname\relax\def\natexlab#1{#1}\fi
\expandafter\ifx\csname bibnamefont\endcsname\relax
  \def\bibnamefont#1{#1}\fi
\expandafter\ifx\csname bibfnamefont\endcsname\relax
  \def\bibfnamefont#1{#1}\fi
\expandafter\ifx\csname citenamefont\endcsname\relax
  \def\citenamefont#1{#1}\fi
\expandafter\ifx\csname url\endcsname\relax
  \def\url#1{\texttt{#1}}\fi
\expandafter\ifx\csname urlprefix\endcsname\relax\def\urlprefix{URL }\fi
\providecommand{\bibinfo}[2]{#2}
\providecommand{\eprint}[2][]{\url{#2}}

\bibitem[{\citenamefont{Tilley and Tilley}(1990)}]{Tilley-book}
\bibinfo{author}{\bibfnamefont{D.}~\bibnamefont{Tilley}} \bibnamefont{and}
  \bibinfo{author}{\bibfnamefont{J.}~\bibnamefont{Tilley}},
  \emph{\bibinfo{title}{Superfluidity and Superconductivity}}
  (\bibinfo{publisher}{Institute of Physics}, \bibinfo{address}{Bristol},
  \bibinfo{year}{1990}), \bibinfo{edition}{3rd} ed.

\bibitem[{\citenamefont{Donnelly}(1991)}]{Donnelly-book}
\bibinfo{author}{\bibfnamefont{R.~J.} \bibnamefont{Donnelly}},
  \emph{\bibinfo{title}{Quantized Vortices in Helium {II}}}
  (\bibinfo{publisher}{Cambridge University Press},
  \bibinfo{address}{Cambridge, England}, \bibinfo{year}{1991}).

\bibitem[{\citenamefont{Vinen and Niemela}(2002)}]{Vinen-2002-JLTP}
\bibinfo{author}{\bibfnamefont{W.~F.} \bibnamefont{Vinen}} \bibnamefont{and}
  \bibinfo{author}{\bibfnamefont{J.~J.} \bibnamefont{Niemela}},
  \bibinfo{journal}{J. Low Temp. Phys.} \textbf{\bibinfo{volume}{128}},
  \bibinfo{pages}{167} (\bibinfo{year}{2002}).

\bibitem[{\citenamefont{Vinen}(1957)}]{Vinen-1957-PRS}
\bibinfo{author}{\bibfnamefont{W.~F.} \bibnamefont{Vinen}},
  \bibinfo{journal}{Proc. Roy. Soc. A} \textbf{\bibinfo{volume}{240}},
  \bibinfo{pages}{114} (\bibinfo{year}{1957}).

\bibitem[{\citenamefont{Stalp et~al.}(1999)\citenamefont{Stalp, Skrbek, and
  Donnelly}}]{Stalp-1999-PRL}
\bibinfo{author}{\bibfnamefont{S.~R.} \bibnamefont{Stalp}},
  \bibinfo{author}{\bibfnamefont{L.}~\bibnamefont{Skrbek}}, \bibnamefont{and}
  \bibinfo{author}{\bibfnamefont{R.~J.} \bibnamefont{Donnelly}},
  \bibinfo{journal}{Phys. Rev. Lett.} \textbf{\bibinfo{volume}{82}},
  \bibinfo{pages}{4831} (\bibinfo{year}{1999}).

\bibitem[{\citenamefont{Maurer and Tabeling}(1998)}]{Maurer-1998-EPL}
\bibinfo{author}{\bibfnamefont{J.}~\bibnamefont{Maurer}} \bibnamefont{and}
  \bibinfo{author}{\bibfnamefont{P.}~\bibnamefont{Tabeling}},
  \bibinfo{journal}{Europhys. Lett.} \textbf{\bibinfo{volume}{43}},
  \bibinfo{pages}{29} (\bibinfo{year}{1998}).

\bibitem[{\citenamefont{Donnelly and Barenghi}(1998)}]{Donnelly-1998-JPCRD}
\bibinfo{author}{\bibfnamefont{R.~J.} \bibnamefont{Donnelly}} \bibnamefont{and}
  \bibinfo{author}{\bibfnamefont{C.~F.} \bibnamefont{Barenghi}},
  \bibinfo{journal}{J. Phys. Chem. Ref. Data} \textbf{\bibinfo{volume}{27}},
  \bibinfo{pages}{1217} (\bibinfo{year}{1998}).

\bibitem[{\citenamefont{Sreenivasan and Donnelly}(2001)}]{Sreenivasan-2001-AAM}
\bibinfo{author}{\bibfnamefont{K.~R.} \bibnamefont{Sreenivasan}}
  \bibnamefont{and} \bibinfo{author}{\bibfnamefont{R.~J.}
  \bibnamefont{Donnelly}}, \bibinfo{journal}{Adv. Appl. Mech.}
  \textbf{\bibinfo{volume}{37}}, \bibinfo{pages}{239} (\bibinfo{year}{2001}).

\bibitem[{\citenamefont{Niemela and Sreenivasan}(2006)}]{Niemela-2006-JLTP}
\bibinfo{author}{\bibfnamefont{J.~J.} \bibnamefont{Niemela}} \bibnamefont{and}
  \bibinfo{author}{\bibfnamefont{K.~R.} \bibnamefont{Sreenivasan}},
  \bibinfo{journal}{J. Low. Temp. Phys.} \textbf{\bibinfo{volume}{143}},
  \bibinfo{pages}{163} (\bibinfo{year}{2006}).

\bibitem[{\citenamefont{Vinen}(2000)}]{Vinen-2000-PRB}
\bibinfo{author}{\bibfnamefont{W.~F.} \bibnamefont{Vinen}},
  \bibinfo{journal}{Phys. Rev. B} \textbf{\bibinfo{volume}{61}},
  \bibinfo{pages}{1410} (\bibinfo{year}{2000}).

\bibitem[{\citenamefont{Barenghi et~al.}(1997)\citenamefont{Barenghi, Samuels,
  Bauer, and Donnelly}}]{Barenghi-1997-PF}
\bibinfo{author}{\bibfnamefont{C.~F.} \bibnamefont{Barenghi}},
  \bibinfo{author}{\bibfnamefont{D.~C.} \bibnamefont{Samuels}},
  \bibinfo{author}{\bibfnamefont{G.~H.} \bibnamefont{Bauer}}, \bibnamefont{and}
  \bibinfo{author}{\bibfnamefont{R.~J.} \bibnamefont{Donnelly}},
  \bibinfo{journal}{Phys. Fluids} \textbf{\bibinfo{volume}{9}},
  \bibinfo{pages}{2631} (\bibinfo{year}{1997}).

\bibitem[{\citenamefont{Bou{\'e} et~al.}(2013)\citenamefont{Bou{\'e}, L'vov,
  Pomyalov, and Procaccia}}]{Boue-2013-PRL}
\bibinfo{author}{\bibfnamefont{L.}~\bibnamefont{Bou{\'e}}},
  \bibinfo{author}{\bibfnamefont{V.}~\bibnamefont{L'vov}},
  \bibinfo{author}{\bibfnamefont{A.}~\bibnamefont{Pomyalov}}, \bibnamefont{and}
  \bibinfo{author}{\bibfnamefont{I.}~\bibnamefont{Procaccia}},
  \bibinfo{journal}{Phys. Rev. Lett.} \textbf{\bibinfo{volume}{110}},
  \bibinfo{pages}{014502} (\bibinfo{year}{2013}).

\bibitem[{\citenamefont{Comte-Bellot and Corrsin}(1966)}]{Comte-1966-JFM}
\bibinfo{author}{\bibfnamefont{G.}~\bibnamefont{Comte-Bellot}}
  \bibnamefont{and} \bibinfo{author}{\bibfnamefont{S.}~\bibnamefont{Corrsin}},
  \bibinfo{journal}{J. Fluid Mech.} \textbf{\bibinfo{volume}{25}},
  \bibinfo{pages}{657} (\bibinfo{year}{1966}).

\bibitem[{\citenamefont{Tennekes and Lumley}(1972)}]{Tennekes-1972-B}
\bibinfo{author}{\bibfnamefont{H.}~\bibnamefont{Tennekes}} \bibnamefont{and}
  \bibinfo{author}{\bibfnamefont{J.~L.} \bibnamefont{Lumley}},
  \emph{\bibinfo{title}{A First Course in Turbulence}} (\bibinfo{publisher}{The
  MIT Press}, \bibinfo{address}{Boston, United States}, \bibinfo{year}{1972}).

\bibitem[{\citenamefont{Sinhuber et~al.}(2015)\citenamefont{Sinhuber,
  Bodenschatz, and Bewley}}]{Sinhuber-2015-PRL}
\bibinfo{author}{\bibfnamefont{M.}~\bibnamefont{Sinhuber}},
  \bibinfo{author}{\bibfnamefont{E.}~\bibnamefont{Bodenschatz}},
  \bibnamefont{and} \bibinfo{author}{\bibfnamefont{G.~P.}
  \bibnamefont{Bewley}}, \bibinfo{journal}{Phys. Rev. Lett.}
  \textbf{\bibinfo{volume}{114}}, \bibinfo{pages}{034501}
  (\bibinfo{year}{2015}).

\bibitem[{\citenamefont{Varga et~al.}(2018)\citenamefont{Varga, Gao, Guo, and
  Skrbek}}]{Varga-2018-PRF}
\bibinfo{author}{\bibfnamefont{E.}~\bibnamefont{Varga}},
  \bibinfo{author}{\bibfnamefont{J.}~\bibnamefont{Gao}},
  \bibinfo{author}{\bibfnamefont{W.}~\bibnamefont{Guo}}, \bibnamefont{and}
  \bibinfo{author}{\bibfnamefont{L.}~\bibnamefont{Skrbek}},
  \bibinfo{journal}{Phys. Rev. Fluids} \textbf{\bibinfo{volume}{3}},
  \bibinfo{pages}{094601} (\bibinfo{year}{2018}).

\bibitem[{\citenamefont{Gao et~al.}(2015)\citenamefont{Gao, Marakov, Guo,
  Pawlowski, Van~Sciver, Ihas, McKinsey, and Vinen}}]{Gao-2015-RSI}
\bibinfo{author}{\bibfnamefont{J.}~\bibnamefont{Gao}},
  \bibinfo{author}{\bibfnamefont{A.}~\bibnamefont{Marakov}},
  \bibinfo{author}{\bibfnamefont{W.}~\bibnamefont{Guo}},
  \bibinfo{author}{\bibfnamefont{B.~T.} \bibnamefont{Pawlowski}},
  \bibinfo{author}{\bibfnamefont{S.~W.} \bibnamefont{Van~Sciver}},
  \bibinfo{author}{\bibfnamefont{G.~G.} \bibnamefont{Ihas}},
  \bibinfo{author}{\bibfnamefont{D.~N.} \bibnamefont{McKinsey}},
  \bibnamefont{and} \bibinfo{author}{\bibfnamefont{W.~F.} \bibnamefont{Vinen}},
  \bibinfo{journal}{Rev. Sci. Instrum.} \textbf{\bibinfo{volume}{86}},
  \bibinfo{pages}{093904} (\bibinfo{year}{2015}).

\bibitem[{\citenamefont{Marakov et~al.}(2015)\citenamefont{Marakov, Gao, Guo,
  Van~Sciver, Ihas, McKinsey, and Vinen}}]{Marakov-2015-PRB}
\bibinfo{author}{\bibfnamefont{A.}~\bibnamefont{Marakov}},
  \bibinfo{author}{\bibfnamefont{J.}~\bibnamefont{Gao}},
  \bibinfo{author}{\bibfnamefont{W.}~\bibnamefont{Guo}},
  \bibinfo{author}{\bibfnamefont{S.~W.} \bibnamefont{Van~Sciver}},
  \bibinfo{author}{\bibfnamefont{G.~G.} \bibnamefont{Ihas}},
  \bibinfo{author}{\bibfnamefont{D.~N.} \bibnamefont{McKinsey}},
  \bibnamefont{and} \bibinfo{author}{\bibfnamefont{W.~F.} \bibnamefont{Vinen}},
  \bibinfo{journal}{Phys. Rev. B} \textbf{\bibinfo{volume}{91}},
  \bibinfo{pages}{094503} (\bibinfo{year}{2015}).

\bibitem[{\citenamefont{Gao et~al.}(2016{\natexlab{a}})\citenamefont{Gao, Guo,
  L'vov, Pomyalov, Skrbek, Varga, and Vinen}}]{Gao-2016-JETP}
\bibinfo{author}{\bibfnamefont{J.}~\bibnamefont{Gao}},
  \bibinfo{author}{\bibfnamefont{W.}~\bibnamefont{Guo}},
  \bibinfo{author}{\bibfnamefont{V.~S.} \bibnamefont{L'vov}},
  \bibinfo{author}{\bibfnamefont{A.}~\bibnamefont{Pomyalov}},
  \bibinfo{author}{\bibfnamefont{L.}~\bibnamefont{Skrbek}},
  \bibinfo{author}{\bibfnamefont{E.}~\bibnamefont{Varga}}, \bibnamefont{and}
  \bibinfo{author}{\bibfnamefont{W.~F.} \bibnamefont{Vinen}},
  \bibinfo{journal}{JETP Lett.} \textbf{\bibinfo{volume}{103}},
  \bibinfo{pages}{648} (\bibinfo{year}{2016}{\natexlab{a}}).

\bibitem[{\citenamefont{Gao et~al.}(2016{\natexlab{b}})\citenamefont{Gao, Guo,
  and Vinen}}]{Gao-2016-PRB}
\bibinfo{author}{\bibfnamefont{J.}~\bibnamefont{Gao}},
  \bibinfo{author}{\bibfnamefont{W.}~\bibnamefont{Guo}}, \bibnamefont{and}
  \bibinfo{author}{\bibfnamefont{W.~F.} \bibnamefont{Vinen}},
  \bibinfo{journal}{Phys. Rev. B} \textbf{\bibinfo{volume}{94}},
  \bibinfo{pages}{094502} (\bibinfo{year}{2016}{\natexlab{b}}).

\bibitem[{\citenamefont{Gao et~al.}(2017{\natexlab{a}})\citenamefont{Gao,
  Varga, Guo, and Vinen}}]{Gao-2017-PRB}
\bibinfo{author}{\bibfnamefont{J.}~\bibnamefont{Gao}},
  \bibinfo{author}{\bibfnamefont{E.}~\bibnamefont{Varga}},
  \bibinfo{author}{\bibfnamefont{W.}~\bibnamefont{Guo}}, \bibnamefont{and}
  \bibinfo{author}{\bibfnamefont{W.~F.} \bibnamefont{Vinen}},
  \bibinfo{journal}{Phys. Rev. B} \textbf{\bibinfo{volume}{96}},
  \bibinfo{pages}{094511} (\bibinfo{year}{2017}{\natexlab{a}}).

\bibitem[{\citenamefont{Gao et~al.}(2017{\natexlab{b}})\citenamefont{Gao,
  Varga, Guo, and Vinen}}]{Gao-2017-JLTP}
\bibinfo{author}{\bibfnamefont{J.}~\bibnamefont{Gao}},
  \bibinfo{author}{\bibfnamefont{E.}~\bibnamefont{Varga}},
  \bibinfo{author}{\bibfnamefont{W.}~\bibnamefont{Guo}}, \bibnamefont{and}
  \bibinfo{author}{\bibfnamefont{W.~F.} \bibnamefont{Vinen}},
  \bibinfo{journal}{J. Low Temp. Phys.} \textbf{\bibinfo{volume}{187}},
  \bibinfo{pages}{490} (\bibinfo{year}{2017}{\natexlab{b}}).

\bibitem[{\citenamefont{Gao et~al.}(2018)\citenamefont{Gao, Guo, Yui, Tsubota,
  and Vinen}}]{Gao-2018-PRB}
\bibinfo{author}{\bibfnamefont{J.}~\bibnamefont{Gao}},
  \bibinfo{author}{\bibfnamefont{W.}~\bibnamefont{Guo}},
  \bibinfo{author}{\bibfnamefont{S.}~\bibnamefont{Yui}},
  \bibinfo{author}{\bibfnamefont{M.}~\bibnamefont{Tsubota}}, \bibnamefont{and}
  \bibinfo{author}{\bibfnamefont{W.~F.} \bibnamefont{Vinen}},
  \bibinfo{journal}{Phys. Rev. B} \textbf{\bibinfo{volume}{97}},
  \bibinfo{pages}{184518} (\bibinfo{year}{2018}).

\bibitem[{\citenamefont{Bao et~al.}(2018)\citenamefont{Bao, Guo, L'vov, and
  Pomyalov}}]{Bao-2018-PRB}
\bibinfo{author}{\bibfnamefont{S.}~\bibnamefont{Bao}},
  \bibinfo{author}{\bibfnamefont{W.}~\bibnamefont{Guo}},
  \bibinfo{author}{\bibfnamefont{V.~S.} \bibnamefont{L'vov}}, \bibnamefont{and}
  \bibinfo{author}{\bibfnamefont{A.}~\bibnamefont{Pomyalov}},
  \bibinfo{journal}{Phys. Rev. B} \textbf{\bibinfo{volume}{98}},
  \bibinfo{pages}{174509} (\bibinfo{year}{2018}).

\bibitem[{\citenamefont{Guo}(2019)}]{Guo-2019-JLTP}
\bibinfo{author}{\bibfnamefont{W.}~\bibnamefont{Guo}}, \bibinfo{journal}{J. Low
  Temp. Phys.} \textbf{\bibinfo{volume}{196}}, \bibinfo{pages}{60}
  (\bibinfo{year}{2019}).

\bibitem[{\citenamefont{Mastracci and
  Guo}(2018{\natexlab{a}})}]{Mastracci-2018-RSI}
\bibinfo{author}{\bibfnamefont{B.}~\bibnamefont{Mastracci}} \bibnamefont{and}
  \bibinfo{author}{\bibfnamefont{W.}~\bibnamefont{Guo}}, \bibinfo{journal}{Rev.
  Sci. Instrum.} \textbf{\bibinfo{volume}{89}}, \bibinfo{pages}{015107}
  (\bibinfo{year}{2018}{\natexlab{a}}).

\bibitem[{\citenamefont{Mastracci et~al.}(2017)\citenamefont{Mastracci, Takada,
  and Guo}}]{Mastracci-2017-JLTP}
\bibinfo{author}{\bibfnamefont{B.}~\bibnamefont{Mastracci}},
  \bibinfo{author}{\bibfnamefont{S.}~\bibnamefont{Takada}}, \bibnamefont{and}
  \bibinfo{author}{\bibfnamefont{W.}~\bibnamefont{Guo}}, \bibinfo{journal}{J.
  Low Temp. Phys.} \textbf{\bibinfo{volume}{187}}, \bibinfo{pages}{446}
  (\bibinfo{year}{2017}).

\bibitem[{\citenamefont{Landau and Lifshitz}(1987)}]{Landau-book}
\bibinfo{author}{\bibfnamefont{L.}~\bibnamefont{Landau}} \bibnamefont{and}
  \bibinfo{author}{\bibfnamefont{E.}~\bibnamefont{Lifshitz}},
  \emph{\bibinfo{title}{Fluid mechanics}} (\bibinfo{publisher}{Pergamon Press},
  \bibinfo{address}{Oxford, England}, \bibinfo{year}{1987}),
  \bibinfo{edition}{2nd} ed.

\bibitem[{\citenamefont{Mastracci and
  Guo}(2018{\natexlab{b}})}]{Mastracci-2018-PRF}
\bibinfo{author}{\bibfnamefont{B.}~\bibnamefont{Mastracci}} \bibnamefont{and}
  \bibinfo{author}{\bibfnamefont{W.}~\bibnamefont{Guo}},
  \bibinfo{journal}{Phys. Rev. Fluids} \textbf{\bibinfo{volume}{3}},
  \bibinfo{pages}{063304} (\bibinfo{year}{2018}{\natexlab{b}}).

\bibitem[{\citenamefont{Mastracci and Guo}(2019)}]{Mastracci-2019-PRF}
\bibinfo{author}{\bibfnamefont{B.}~\bibnamefont{Mastracci}} \bibnamefont{and}
  \bibinfo{author}{\bibfnamefont{W.}~\bibnamefont{Guo}},
  \bibinfo{journal}{Phys. Rev. Fluids} \textbf{\bibinfo{volume}{4}},
  \bibinfo{pages}{023301} (\bibinfo{year}{2019}).

\bibitem[{\citenamefont{Mastracci et~al.}(2019)\citenamefont{Mastracci, Bao,
  Guo, and Vinen}}]{Mastracci-2019-PRF-2}
\bibinfo{author}{\bibfnamefont{B.}~\bibnamefont{Mastracci}},
  \bibinfo{author}{\bibfnamefont{S.}~\bibnamefont{Bao}},
  \bibinfo{author}{\bibfnamefont{W.}~\bibnamefont{Guo}}, \bibnamefont{and}
  \bibinfo{author}{\bibfnamefont{W.~F.} \bibnamefont{Vinen}},
  \bibinfo{journal}{Phys. Rev. Fluids} \textbf{\bibinfo{volume}{4}},
  \bibinfo{pages}{083305} (\bibinfo{year}{2019}).

\bibitem[{\citenamefont{Fernando and De~Silva}(1993)}]{Fernando-1993-PoF}
\bibinfo{author}{\bibfnamefont{H.}~\bibnamefont{Fernando}} \bibnamefont{and}
  \bibinfo{author}{\bibfnamefont{I.}~\bibnamefont{De~Silva}},
  \bibinfo{journal}{Phys. Fluids. A-Fluid} \textbf{\bibinfo{volume}{5}},
  \bibinfo{pages}{1849} (\bibinfo{year}{1993}).

\bibitem[{\citenamefont{Honey et~al.}(2014)\citenamefont{Honey, Hershberger,
  Donnelly, and Bolster}}]{Honey-2014-PRE}
\bibinfo{author}{\bibfnamefont{R.~E.} \bibnamefont{Honey}},
  \bibinfo{author}{\bibfnamefont{R.}~\bibnamefont{Hershberger}},
  \bibinfo{author}{\bibfnamefont{R.~J.} \bibnamefont{Donnelly}},
  \bibnamefont{and} \bibinfo{author}{\bibfnamefont{D.}~\bibnamefont{Bolster}},
  \bibinfo{journal}{Phys. Rev. E} \textbf{\bibinfo{volume}{89}},
  \bibinfo{pages}{053016} (\bibinfo{year}{2014}).

\bibitem[{\citenamefont{Fonda et~al.}(2016)\citenamefont{Fonda, Sreenivasan,
  and Lathrop}}]{Fonda-2016-RSI}
\bibinfo{author}{\bibfnamefont{E.}~\bibnamefont{Fonda}},
  \bibinfo{author}{\bibfnamefont{K.~R.} \bibnamefont{Sreenivasan}},
  \bibnamefont{and} \bibinfo{author}{\bibfnamefont{D.~P.}
  \bibnamefont{Lathrop}}, \bibinfo{journal}{Rev. Sci. Instrum.}
  \textbf{\bibinfo{volume}{87}}, \bibinfo{pages}{025106}
  (\bibinfo{year}{2016}).

\bibitem[{\citenamefont{Sbalzarini and
  Koumoutsakos}(2005)}]{Sbalzarini-2005-JSB}
\bibinfo{author}{\bibfnamefont{I.~F.} \bibnamefont{Sbalzarini}}
  \bibnamefont{and}
  \bibinfo{author}{\bibfnamefont{P.}~\bibnamefont{Koumoutsakos}},
  \bibinfo{journal}{J. Struct. Biol.} \textbf{\bibinfo{volume}{151}},
  \bibinfo{pages}{182} (\bibinfo{year}{2005}).

\bibitem[{\citenamefont{Skrbek and Stalp}(2000)}]{Skrbek-2000-PoF}
\bibinfo{author}{\bibfnamefont{L.}~\bibnamefont{Skrbek}} \bibnamefont{and}
  \bibinfo{author}{\bibfnamefont{S.~R.} \bibnamefont{Stalp}},
  \bibinfo{journal}{Phys. Fluids} \textbf{\bibinfo{volume}{12}},
  \bibinfo{pages}{1997} (\bibinfo{year}{2000}).

\bibitem[{\citenamefont{Davidson}(2004)}]{Davidson-2004-book}
\bibinfo{author}{\bibfnamefont{P.~A.} \bibnamefont{Davidson}},
  \emph{\bibinfo{title}{Turbulence: An Introduction for Scientists and
  Engineers}} (\bibinfo{publisher}{Oxford University Press},
  \bibinfo{address}{Oxford, United Kingdom}, \bibinfo{year}{2004}).

\bibitem[{\citenamefont{Stolovitzky and
  Sreenivasan}(1993)}]{Stolovitzky-1993-PRE}
\bibinfo{author}{\bibfnamefont{G.}~\bibnamefont{Stolovitzky}} \bibnamefont{and}
  \bibinfo{author}{\bibfnamefont{K.~R.} \bibnamefont{Sreenivasan}},
  \bibinfo{journal}{Phys. Rev. E} \textbf{\bibinfo{volume}{48}},
  \bibinfo{pages}{R33} (\bibinfo{year}{1993}).

\bibitem[{\citenamefont{Benzi et~al.}(1995)\citenamefont{Benzi, Ciliberto,
  Baudet, and Chavarria}}]{Benzi-1995-PhysicaD}
\bibinfo{author}{\bibfnamefont{R.}~\bibnamefont{Benzi}},
  \bibinfo{author}{\bibfnamefont{S.}~\bibnamefont{Ciliberto}},
  \bibinfo{author}{\bibfnamefont{C.}~\bibnamefont{Baudet}}, \bibnamefont{and}
  \bibinfo{author}{\bibfnamefont{G.~R.} \bibnamefont{Chavarria}},
  \bibinfo{journal}{Physica D} \textbf{\bibinfo{volume}{80}},
  \bibinfo{pages}{385} (\bibinfo{year}{1995}).

\bibitem[{\citenamefont{Frisch and Kolmogorov}(1995)}]{Frisch-1995-B}
\bibinfo{author}{\bibfnamefont{U.}~\bibnamefont{Frisch}} \bibnamefont{and}
  \bibinfo{author}{\bibfnamefont{A.~N.} \bibnamefont{Kolmogorov}},
  \emph{\bibinfo{title}{Turbulence: the legacy of A.N. Kolmogorov}}
  (\bibinfo{publisher}{Cambridge university press},
  \bibinfo{address}{Cambridge, United Kingdom}, \bibinfo{year}{1995}).

\bibitem[{\citenamefont{Hinze}(1975)}]{Hinze-1975-B}
\bibinfo{author}{\bibfnamefont{J.}~\bibnamefont{Hinze}},
  \emph{\bibinfo{title}{Turbulence}} (\bibinfo{publisher}{MacGraw Hill,
  New-York}, \bibinfo{year}{1975}), \bibinfo{edition}{2nd} ed.

\bibitem[{\citenamefont{She and Leveque}(1994)}]{She-1994-PRL}
\bibinfo{author}{\bibfnamefont{Z.-S.} \bibnamefont{She}} \bibnamefont{and}
  \bibinfo{author}{\bibfnamefont{E.}~\bibnamefont{Leveque}},
  \bibinfo{journal}{Phys. Rev. Lett.} \textbf{\bibinfo{volume}{72}},
  \bibinfo{pages}{336} (\bibinfo{year}{1994}).

\bibitem[{\citenamefont{Benzi et~al.}(1993{\natexlab{a}})\citenamefont{Benzi,
  Ciliberto, Tripiccione, Baudet, Massaioli, and Succi}}]{Benzi-1993-PRE}
\bibinfo{author}{\bibfnamefont{R.}~\bibnamefont{Benzi}},
  \bibinfo{author}{\bibfnamefont{S.}~\bibnamefont{Ciliberto}},
  \bibinfo{author}{\bibfnamefont{R.}~\bibnamefont{Tripiccione}},
  \bibinfo{author}{\bibfnamefont{C.}~\bibnamefont{Baudet}},
  \bibinfo{author}{\bibfnamefont{F.}~\bibnamefont{Massaioli}},
  \bibnamefont{and} \bibinfo{author}{\bibfnamefont{S.}~\bibnamefont{Succi}},
  \bibinfo{journal}{Phys. Rev. E} \textbf{\bibinfo{volume}{48}},
  \bibinfo{pages}{R29} (\bibinfo{year}{1993}{\natexlab{a}}).

\bibitem[{\citenamefont{Salort et~al.}(2011)\citenamefont{Salort, Chabaud,
  L{\'e}v{\^e}que, and Roche}}]{Salort-2011-JoP}
\bibinfo{author}{\bibfnamefont{J.}~\bibnamefont{Salort}},
  \bibinfo{author}{\bibfnamefont{B.}~\bibnamefont{Chabaud}},
  \bibinfo{author}{\bibfnamefont{E.}~\bibnamefont{L{\'e}v{\^e}que}},
  \bibnamefont{and} \bibinfo{author}{\bibfnamefont{P.-E.} \bibnamefont{Roche}},
  in \emph{\bibinfo{booktitle}{J. Phys.: Conf. Ser.}}
  (\bibinfo{organization}{IOP Publishing}, \bibinfo{year}{2011}), vol.
  \bibinfo{volume}{318}, p. \bibinfo{pages}{042014}.

\bibitem[{\citenamefont{Benzi et~al.}(1993{\natexlab{b}})\citenamefont{Benzi,
  Ciliberto, Baudet, Chavarria, and Tripiccione}}]{Benzi-1993-EPL}
\bibinfo{author}{\bibfnamefont{R.}~\bibnamefont{Benzi}},
  \bibinfo{author}{\bibfnamefont{S.}~\bibnamefont{Ciliberto}},
  \bibinfo{author}{\bibfnamefont{C.}~\bibnamefont{Baudet}},
  \bibinfo{author}{\bibfnamefont{G.~R.} \bibnamefont{Chavarria}},
  \bibnamefont{and}
  \bibinfo{author}{\bibfnamefont{R.}~\bibnamefont{Tripiccione}},
  \bibinfo{journal}{Europhys. Lett.} \textbf{\bibinfo{volume}{24}},
  \bibinfo{pages}{275} (\bibinfo{year}{1993}{\natexlab{b}}).

\bibitem[{\citenamefont{Dubrulle}(1994)}]{Dubrulle-1994-PRL}
\bibinfo{author}{\bibfnamefont{B.}~\bibnamefont{Dubrulle}},
  \bibinfo{journal}{Phys. Rev. Lett.} \textbf{\bibinfo{volume}{73}},
  \bibinfo{pages}{959} (\bibinfo{year}{1994}).

\bibitem[{\citenamefont{Stolovitzky et~al.}(1993)\citenamefont{Stolovitzky,
  Sreenivasan, and Juneja}}]{Stolovitzky-1993-PRE1}
\bibinfo{author}{\bibfnamefont{G.}~\bibnamefont{Stolovitzky}},
  \bibinfo{author}{\bibfnamefont{K.~R.} \bibnamefont{Sreenivasan}},
  \bibnamefont{and} \bibinfo{author}{\bibfnamefont{A.}~\bibnamefont{Juneja}},
  \bibinfo{journal}{Phys. Rev. E} \textbf{\bibinfo{volume}{48}},
  \bibinfo{pages}{R3217} (\bibinfo{year}{1993}).

\bibitem[{\citenamefont{Rusaouen et~al.}(2017)\citenamefont{Rusaouen, Chabaud,
  Salort, and Roche}}]{Rusaouen-2017-PoF}
\bibinfo{author}{\bibfnamefont{E.}~\bibnamefont{Rusaouen}},
  \bibinfo{author}{\bibfnamefont{B.}~\bibnamefont{Chabaud}},
  \bibinfo{author}{\bibfnamefont{J.}~\bibnamefont{Salort}}, \bibnamefont{and}
  \bibinfo{author}{\bibfnamefont{P.-E.} \bibnamefont{Roche}},
  \bibinfo{journal}{Phys. Fluids} \textbf{\bibinfo{volume}{29}},
  \bibinfo{pages}{105108} (\bibinfo{year}{2017}).

\bibitem[{\citenamefont{Bewley et~al.}(2006)\citenamefont{Bewley, Lathrop, and
  Sreenivasan}}]{Bewley-2006-Nature}
\bibinfo{author}{\bibfnamefont{G.~P.} \bibnamefont{Bewley}},
  \bibinfo{author}{\bibfnamefont{D.~P.} \bibnamefont{Lathrop}},
  \bibnamefont{and} \bibinfo{author}{\bibfnamefont{K.~R.}
  \bibnamefont{Sreenivasan}}, \bibinfo{journal}{Nature}
  \textbf{\bibinfo{volume}{441}}, \bibinfo{pages}{588} (\bibinfo{year}{2006}).

\bibitem[{\citenamefont{Zhang and Van~Sciver}(2005)}]{Zhang-2005-NPhys}
\bibinfo{author}{\bibfnamefont{T.}~\bibnamefont{Zhang}} \bibnamefont{and}
  \bibinfo{author}{\bibfnamefont{S.~W.} \bibnamefont{Van~Sciver}},
  \bibinfo{journal}{Nature Phys.} \textbf{\bibinfo{volume}{1}},
  \bibinfo{pages}{36} (\bibinfo{year}{2005}).

\bibitem[{\citenamefont{Idowu et~al.}(2000)\citenamefont{Idowu, Willis,
  Barenghi, and Samuels}}]{Idowu-2000-PRB}
\bibinfo{author}{\bibfnamefont{O.~C.} \bibnamefont{Idowu}},
  \bibinfo{author}{\bibfnamefont{A.}~\bibnamefont{Willis}},
  \bibinfo{author}{\bibfnamefont{C.~F.} \bibnamefont{Barenghi}},
  \bibnamefont{and} \bibinfo{author}{\bibfnamefont{D.~C.}
  \bibnamefont{Samuels}}, \bibinfo{journal}{Phys. Rev. B}
  \textbf{\bibinfo{volume}{62}}, \bibinfo{pages}{3409} (\bibinfo{year}{2000}).

\bibitem[{\citenamefont{Yui et~al.}(2019)\citenamefont{Yui, Kobayashi, Tsubota,
  and Guo}}]{Yui-2020-PRL}
\bibinfo{author}{\bibfnamefont{S.}~\bibnamefont{Yui}},
  \bibinfo{author}{\bibfnamefont{H.}~\bibnamefont{Kobayashi}},
  \bibinfo{author}{\bibfnamefont{M.}~\bibnamefont{Tsubota}}, \bibnamefont{and}
  \bibinfo{author}{\bibfnamefont{W.}~\bibnamefont{Guo}} (\bibinfo{year}{2019}),
  \eprint{1911.01628}.

\bibitem[{\citenamefont{Kivotides et~al.}(2008)\citenamefont{Kivotides,
  Barenghi, and Sergeev}}]{Kivotides-2008-PRB}
\bibinfo{author}{\bibfnamefont{D.}~\bibnamefont{Kivotides}},
  \bibinfo{author}{\bibfnamefont{C.~F.} \bibnamefont{Barenghi}},
  \bibnamefont{and} \bibinfo{author}{\bibfnamefont{Y.~A.}
  \bibnamefont{Sergeev}}, \bibinfo{journal}{Phys. Rev. B}
  \textbf{\bibinfo{volume}{77}}, \bibinfo{pages}{014527}
  (\bibinfo{year}{2008}).

\bibitem[{\citenamefont{Kivotides}(2008)}]{Kivotides-2008-PRB-2}
\bibinfo{author}{\bibfnamefont{D.}~\bibnamefont{Kivotides}},
  \bibinfo{journal}{Phys. Rev. B} \textbf{\bibinfo{volume}{77}},
  \bibinfo{pages}{174508} (\bibinfo{year}{2008}).

\bibitem[{\citenamefont{Arneodo et~al.}(1996)\citenamefont{Arneodo, Baudet,
  Belin, Benzi, Castaing, Chabaud, Chavarria, Ciliberto, Camussi, Chilla
  et~al.}}]{Arneodo-1996-EPL}
\bibinfo{author}{\bibfnamefont{A.}~\bibnamefont{Arneodo}},
  \bibinfo{author}{\bibfnamefont{C.}~\bibnamefont{Baudet}},
  \bibinfo{author}{\bibfnamefont{F.}~\bibnamefont{Belin}},
  \bibinfo{author}{\bibfnamefont{R.}~\bibnamefont{Benzi}},
  \bibinfo{author}{\bibfnamefont{B.}~\bibnamefont{Castaing}},
  \bibinfo{author}{\bibfnamefont{B.}~\bibnamefont{Chabaud}},
  \bibinfo{author}{\bibfnamefont{R.}~\bibnamefont{Chavarria}},
  \bibinfo{author}{\bibfnamefont{S.}~\bibnamefont{Ciliberto}},
  \bibinfo{author}{\bibfnamefont{R.}~\bibnamefont{Camussi}},
  \bibinfo{author}{\bibfnamefont{F.}~\bibnamefont{Chilla}},
  \bibnamefont{et~al.}, \bibinfo{journal}{Europhys. Lett.}
  \textbf{\bibinfo{volume}{34}}, \bibinfo{pages}{411} (\bibinfo{year}{1996}).

\end{thebibliography}
\end{document}